\documentclass[runningheads]{llncs}

\usepackage{latexsym,amssymb,amsmath,bm}
\usepackage{graphics,graphicx}
\usepackage{ifthen}
\usepackage{ifthen}
\usepackage{xspace}
\usepackage{paralist}  %
\usepackage{multirow}  % from 2015-16 DNA paper with Lila,Steffen
\usepackage{shuffle}   %
\usepackage{environ}   %
\usepackage[colorlinks,citecolor=blue]{hyperref}

\usepackage{boxedminipage}
\usepackage{marginnote} % e.g., enter \NOTE{....} (see defs.txt)
\usepackage{verbatim}  % allows \begin{comment}.....\end{comment}
\usepackage[usenames,dvipsnames]{xcolor}

\def\MINE{0}
\def\DRAFT{0}
\mathchardef\mhyphen="2D

\newcommand\pssi{\par\smallskip\indent}

\newcommand\pssn{\par\smallskip\noindent}
\newcommand\pmsn{\par\medskip\noindent}
\newcommand\pbsn{\par\bigskip\noindent}
\newcommand\pnsi{\par\indent}
\newcommand\pnsn{\par\noindent}

\newcommand\emdef[1]{\textsf{#1}}  % definition
\newcommand\emshort[1]{\textbf{#1}}
\newcommand\emlong[1]{\textit{#1}}

\newcommand{\sk}[1]{\textit{\textbf{\color{blue}[---$>$ #1]}}}

\newcommand{\NOTE}[1]{\marginnote{\footnotesize\color{red}#1}}
\newcommand{\TEMP}[1]{\if\DRAFT1\marginpar{{\color{red}#1}}\fi}
\newcommand{\sklabel}[1]{\label{#1}\if\DRAFT1\NOTE{#1}\fi}

\newlength{\algoindent}   % for Algorithms
\newcommand{\qtim}[1]{\quad\hbox{#1}\quad}

\if\MINE1
\theoremstyle{plain}
\newtheorem{theorem}{Theorem}[]
\newtheorem{proposition}[theorem]{Proposition}
\newtheorem{lemma}[theorem]{Lemma}
\newtheorem{corollary}[theorem]{Corollary}

\theoremstyle{definition}
\newtheorem{definition}[theorem]{Definition}
\newtheorem{example}[theorem]{Example}
\theoremstyle{remark}
\newtheorem{remark}[theorem]{Remark}
\fi

\newcommand{\cE}{\ensuremath{\mathcal{E}}\xspace}

\renewcommand{\phi}{\varphi}
\renewcommand{\epsilon}{\varepsilon}
\newcommand{\sse}{\subseteq}

\newcommand{\es}{\emptyset}
\newcommand{\sm}{-}  %{\setminus}

\newcommand\card[1]{|#1|}
\newcommand{\prob}[1]{\mathrm{P}[#1]}
\newcommand{\N}{\ensuremath{\mathbb{N}}\xspace}
\newcommand{\Ns}[1]{\ensuremath{\N_{#1}}\xspace}

\usepackage{tikz}
\usetikzlibrary{%
  arrows,%
  positioning,%
  decorations.pathmorphing,%
  decorations.pathreplacing,%
  automata,%
}

\newcommand{\ew}{\ensuremath{\bm\varepsilon}\xspace}      % empty word
\newcommand\al{\mathtt{A}}        %alphabet
\newcommand\len[1]{|#1|}       %length of object
\newcommand{\RP}{\ensuremath{\mathsf{RP}}\xspace}

\newcommand\aut{\ensuremath{\bm{a}}\xspace}   %arbitrary automaton
\newcommand\autb{\ensuremath{\bm{b}}\xspace}   % automaton b
\newcommand{\lang}[1]{\ensuremath\mathtt{L}(#1)\xspace} %language accepted by
\newcommand{\tr}{\ensuremath{\bm{t}}\xspace}       % arbitrary transducer
\newcommand\sz[1]{|#1|}       %size of object
\newcommand{\NFA}[0]{\ensuremath{\mathsf{NFA}}\xspace}
\newcommand{\DFA}[0]{\ensuremath{\mathsf{DFA}}\xspace}
\newcommand{\ADFA}[0]{\ensuremath{\mathsf{ADFA}}\xspace}
\newcommand{\BNFA}[0]{\ensuremath{\mathsf{BNFA}}\xspace}

\newcommand\autac{\ensuremath{\mathbf{a}^{\mathrm c}}\xspace}   %complement of DFA automaton
\newcommand{\zerone}[0]{\ensuremath{(0,1)}\xspace}
\newcommand{\val}[0]{\ensuremath{v}\xspace} %alphabet size

\newcommand{\err}{\ensuremath{\epsilon}\xspace}  %error parameter
\newcommand{\asz}[0]{\ensuremath{s}\xspace} %alphabet size
\newcommand\als{\ensuremath{\al_{\asz}}\xspace}        % alphabet of certain size {0,1,...,s-1}
\newcommand{\dlen}[1]{\ensuremath{\mathop{\rm len}#1}\xspace}   %length
\newcommand{\cnt}{\ensuremath{\mathtt{Cnt}}\xspace}
\newcommand{\icnt}{\ensuremath{\mathtt{I}}\xspace}
\newcommand{\bcnt}{\ensuremath{\mathtt{B}}\xspace}
\newcommand{\upp}{\ensuremath{p\%}\xspace}  %universality percent p
\newcommand{\upar}{\ensuremath{p}\xspace}  %universality parameter
\newcommand{\uparg}{\ensuremath{g}\xspace}  %universality parameter g
\newcommand{\ulm}{\ensuremath{\mathtt{M}}\xspace}  %universal language M
\newcommand{\dom}{\mathop{\rm dom}}
\newcommand{\ev}[1]{\ensuremath{\cE(#1)}\xspace} %expected val
\newcommand{\select}[0]{\stackrel{\$}{\longleftarrow}}
\newcommand{\wrd}[0]{\ensuremath{W}\xspace} % word distr
\newcommand{\wrdm}[0]{\ensuremath{W}^{M}\xspace} % word distr up to M
\newcommand{\wdt}[0]{\ensuremath{T}\xspace} % tractable word distr 
\newcommand{\wdgd}[0]{\ensuremath{\lbdu{\ldla{s,d}}}\xspace} % L word distr s,d
\newcommand{\wdrd}[0]{\ensuremath{\lbdu{\lddi{t,d}}}\xspace} % D word distr t,d
\newcommand{\wdgt}[0]{\ensuremath{\lbdu{\ldla{2,1}}}\xspace} % word distr G_2,1
\newcommand{\wdrt}[0]{\ensuremath{\lbdu{\lddi{2,1}}}\xspace} % word distr R_2,1
\newcommand{\dd}[0]{\ensuremath{D}\xspace} %  distribution d
\newcommand{\uwd}[1]{\ensuremath{\mathrm{U}_{#1}}\xspace} % unif. word distr
\newcommand{\ld}[0]{\ensuremath{N}\xspace} %length distri
\newcommand{\ldun}[1]{\ensuremath{\mathsf{U}_{#1}}\xspace}

\newcommand{\ldla}[1]{\ensuremath{\mathsf{L}_{#1}}\xspace} % Lambert
\newcommand{\lddi}[1]{\ensuremath{\mathsf{D}_{#1}}\xspace} % Dirichlet
\newcommand{\lbdu}[1]{\ensuremath{\langle#1\rangle}\xspace} % length based distr

\newcommand{\unfa}[0]{\textsf{UNIV\_NFA}\xspace}

\newcommand{\ubnfa}[0]{\textsf{UNIV\_BNFA}\xspace}
\newcommand{\umaxlennfa}[0]{\textsf{UNIV\_MAXLEN\_NFA}\xspace}
\newcommand{\adfasubsetnfa}[0]{\textsf{ADFA\_SUBSET\_NFA}\xspace}

\newcommand{\edfai}[0]{\textsf{EMPTY\_DFA}\xspace}

\newcommand{\probd}[1]{\mathsf{prob}_{#1}} % probability based on a distribution

\newcommand\tosscoin{{\mathsf{tossCoin}}\xspace}
\newcommand\selectfin{\mathsf{selectFin}}

\newcommand\uselect{\mathsf{selectUnif}}
\newcommand\uiestim{{\mathsf{UnivIndex}}\xspace}
\newcommand\uiestimml{{\mathsf{UnivIndexMaxLen}}\xspace}
\newcommand\maxlen{{\mathsf{maxLen}}\xspace}
\newcommand\umaxlenalgo{\mathsf{UnivMaxLenNFA}\xspace}
\newcommand\ubalgo{\mathsf{UnivBlockNFA}\xspace}

\newcommand\ualgo{\mathsf{UnivNFA}\xspace}
\newcommand\ualgounary{\mathsf{UnivUnaryNFA}\xspace}
\newcommand\none{\ensuremath{\bot}\xspace}
\newcommand\true{\ensuremath{\mathtt{True}}\xspace}
\newcommand\false{\ensuremath{\mathtt{False}}\xspace}

\begin{document}

\if\MINE1
\begin{center}
%%%%%%%%%%%%%  T I T L E
\textbf{\Large  Approximate NFA Universality and Related %Hard NFA 
Problems Motivated by Information Theory\footnote{Research supported by NSERC, Canada (Discovery Grants of S.K. and of M.M.) and by CMUP   through FCT project UIDB/00144/2021.}}
%\textbf{\Large \upp-Universality and Randomized Approximations of Hard NFA Problems}
%
%%%%%%%%%%%%  AUTHORS
\pbsn
{\large Stavros Konstantinidis$^{1}$,
Mitja Mastnak$^{1}$, 
Nelma Moreira$^{2}$, Rog{\'e}rio Reis$^{2}$}
\end{center}
\pssn
$^{1}$ Department of Mathematics and Computing Science,
Saint Mary's University, Halifax, Nova Scotia, B3H 3C3, 
Canada,
\texttt{s.konstantinidis@smu.ca, mmastnak@cs.smu.ca}
\pssn
$^{2}$  CMUP \& DCC, Faculdade de Ci{\^e}ncias da Universidade do Porto,
Rua do Campo Alegre, 4169â€-007 Porto, Portugal,
\texttt{\{nelma.moreira,rogerio.reis\}@fc.up.pt}

%%%%%%%%%%% ABSTRACT  and K E Y W O R D S
\pbsn
\textbf{Abstract.}

\pbsn
\fi   % end of \if\MINE1

\title{Approximate NFA Universality and Related %Hard NFA 
Problems Motivated by Information Theory\thanks{Research supported by NSERC, Canada (Discovery Grants of S.K. and of M.M.) and by CMUP   through FCT project UIDB/00144/2020.}}

\author{Stavros~Konstantinidis\inst{1} \and Mitja~Mastnak\inst{1} \and Nelma~Moreira\inst{2} \and Rog{\'e}rio~Reis\inst{2}}
\institute{
Saint Mary's University, Halifax, Nova Scotia, Canada,\\
\email{s.konstantinidis@smu.ca, mmastnak@cs.smu.ca}, 
\and
 CMUP \& DM, DCC, Faculdade de
  Ci{\^e}ncias da Universidade do Porto,\\
  Rua do Campo Alegre, 4169-007 Porto, Portugal\\
  \email{\{nelma.moreira,rogerio.reis\}@fc.up.pt}
}
\titlerunning{Approximate NFA Universality and Related %Hard NFA 
Problems}
\authorrunning{S.~Konstantinidis, M.~Mastnak, N.~Moreira, R.~Reis}

\maketitle

\begin{abstract}
In coding and information theory, it is desirable to construct maximal codes that can be either  variable length codes  or error control codes of fixed length. However deciding code maximality  boils down to deciding whether a given NFA is universal, and this is a hard problem (including the case of whether the NFA accepts all words of a fixed length). On the other hand, it is acceptable to know whether a code is `approximately' maximal, which then boils down to whether a given NFA is `approximately' universal. Here we introduce the notion of  a $(1-\err)$-universal automaton and present  polynomial randomized  approximation algorithms to test NFA universality and related hard automata problems, for certain natural probability distributions on the set of words. We also conclude that the randomization aspect is necessary, as  approximate universality remains hard for any fixed polynomially computable~$\err$.
\end{abstract}

\parbox{0.85\textwidth}\qquad
{\color{blue}
\pnsn {\small\textbf{Changes with respect to the 1st version:} (i) In Corollary~\ref{cor:riem}, the bound $O(1)$ for the $M$ of the Dirichlet distribution was incorrect. This bound has been corrected here. (ii) Definition~\ref{def:tractable} of tractable distribution was too restrictive as it required that $M$ is $O\big((\log(1/\err))^k\big)$, for some $k$, but now we see it is necessary that $M$ is $O\big((1/\err)^k\big)$, which is still polynomial wrt $(1/\err)$. (iii) Due to the above change, the PAX algorithm of Section~\ref{sec:PAXforUniv} works only for unary NFAs, so for general NFAs it is necessary to use the PRAX algorithm.}
}

%%%%%%%%%%%%%%%%%%%%%%%%%%%%%%%%%%%%%%%%%%%%%%%%%%%%%%%%%
\section{Introduction}\label{sec:intro}
It is well-known that NFA universality is a PSPACE-hard problem and that block NFA universality (whether an NFA of some fixed length words accepts all the words of that length) is a coNP-hard problem. Here we consider polynomial approximation algorithms for these and related NFA problems by considering the concept of an approximate universal NFA, or block NFA, where for instance 95\% of all words are accepted by the NFA. In general, for some tolerance $\err\in(0,1)$, we assume that we are happy to know that an NFA is at least $(1-\err)$ universal. While approximate universality is still hard, it allows us to consider polynomial randomized algorithms that return an incorrect answer with small probability. Inspired from \cite[pg 72]{MiUp:2017}, we view estimating the universality index of an NFA as the problem of estimating the parameter of some population and then follow the tools of \cite{MiUp:2017} for parameter estimation problems.

Our motivation for defining the concept of approximate universality comes from the problem of generating codes (whether variable length codes, or fixed length error control codes) that are maximal, where on the one hand the question of deciding maximality is hard, but on the other hand it is acceptable to generate codes that are maximal within a tolerance \err, \cite{DudKon:2012,KMR:2018}. 
For infinite languages, we define approximate universality relative to some probability distribution on the set of words. 
This idea is consistent with our interpretation of languages in the context of coding and information theory where words are in fact abstractions of physical network signals or magnetic polarities, \cite{MaRoSi:2001,Jurg:2008}, and the amount of energy they require should not be exponential.

Our work falls under   %\TEMP{Added\\ paragraph}
the general framework of problems about  parameter estimation or approximate counting \cite{MiUp:2017,Gold:2008,ArBa:2009}, however, we are not aware of the application  of this framework in  hard  NFA problems, especially in the case where the NFA accepts an infinite language.

\pnsi\textbf{Main results and  structure of the paper.} 
The next section contains basic notation from formal languages and automata as well concepts of probability distributions on the nonnegative integers, in particular the three distributions:  uniform, Lambert and Dirichlet. The Dirichlet distribution is a good substitute for the `fictitious' uniform distribution on the nonnegative integers \cite{Gol:1970}.
\emshort{Section~\ref{sec:randappr}} discusses what a polynomial randomized approximation (\emdef{PRAX}) algorithm should be for the case of a hard decision problem on NFAs. The necessity for PRAX-like algorithms for NFA universality is demonstrated with (i) the observation that a nonrandomized polynomial approximation (\emdef{PAX}) algorithm might not exist and (ii) the result that $(1-\err)$ approximate block NFA universality  is hard for every \err that is computable within polynomial time. 
\emshort{Section~\ref{sec:worddistr}} is about probability distributions on words over some alphabet $\als=\{0,1,\ldots,s\}$ such that the length sets of these distributions follow the above three distributions on the nonnegatives. 
\emshort{Section~\ref{sec:univ}}  considers whether an  NFA $\aut$ is universal relative to a maximum language \ulm (i.e., whether $\lang{\aut}=\ulm$), and takes the approach that $\ulm$ is the domain of a probability distribution $W$ on the set of words, in which case the universality index $W(\aut)$ of \aut is the probability that a word selected from the distribution $W$ belongs to $\lang{\aut}$. Then, \aut is \upp-universal relative to $W$ if $W(\aut)\ge\upp$. The section closes with two simple random processes about estimating the universality index of NFAs. 
\emshort{Section~\ref{sec:urNFA}} gives PRAX algorithms for three hard NFA problems: \adfasubsetnfa (whether $\lang{\autb}\subseteq\lang{\aut}$ for given NFA \aut and acyclic DFA \autb); \ubnfa (whether $\lang{\aut}=\als^\ell$, for given block NFA \aut of word length $\ell$); and \umaxlennfa (whether $\als^{\le\ell}\subseteq\lang{\aut}$, for given  NFA \aut and  word length $\ell$). 
\emshort{Section~7} defines what a tractable length distribution (on the nonnegatives) is and gives a PRAX algorithm for whether a given NFA is universal relative to any fixed, but arbitrary, tractable word distribution (including the word distributions that are based on the Lambert and Dirichlet length distributions). The section also discusses a  PAX algorithm for universality of unary NFAs relative to a tractable distribution. The last section contains a few concluding results and a short discussion on related hard problems. 
\if\DRAFT1
\pssi\sk{TO BE FINALIZED}
\fi

%%%%%%%%%%%%%%%%%%%%%%%%%%%%%%%%%%%%%%%%%%%%%%%%%%%%%%%%%
\section{Basic Notation and Background Information}\label{sec:notation}
We use the notation \N for the set of  positive integers,  \Ns0 for the nonnegative integers, and $\N^{>x}$ for the positive integers greater than $x$, where $x$ is any real number.
We assume the reader to be familiar with basics of formal languages and finite automata \cite{FLhandbookI,HoMoUl:2001}. Our arbitrary alphabet will be $\als=\{0,1,\ldots,s-1\}$ for some positive  integer $s$. Then, we use the following notation
\pssi
\ew = empty word, \qquad $|w|$ = length of word $w$
\pssi
$\als^\ell$ = all words of length $\ell$,\qquad $\als^{\le\ell}$ = all words of length at most $\ell$
\pssi
$\DFA$ = all  DFAs (deterministic finite automata)
\pssi
$\NFA$ = all  NFAs (nondeterministic finite automata)
\pssi
$\ADFA$ = all acyclic DFAs (accepting finite languages)
\pssi
$\BNFA$ = all block NFAs, that is, NFAs accepting languages of a fixed word length.
\pssi
$\BNFA[s]$ = all block NFAs over the alphabet $\al_s$.
%\pssi
%A \emdef{unary NFA} is an NFA accepting a language over the unary alphabet \{0\}.
\pssi
$\sz\aut$ = the size of the NFA \aut = the number of states plus the number of transitions in \aut.
\pssi$\lang{\aut}$ = the language accepted by the NFA, or DFA, $\aut$.
%\pssn
%The \emdef{density} of a language $L$, \cite{SYZS:1992}, is the function $\dens L:\N_0\to[0,1]$ such that $$\dens L(n)=\frac{|L\cap\als^n|}{|\als^n|}.$$
\pssn\emshort{Notes}: We assume that NFAs have no $\ew$-transitions. It makes no difference in this paper whether a DFA is complete or incomplete.
\pmsn
Next we list some  decision problems about automata that are known to be hard, or easily shown to be hard.
\begin{description}
\item{\unfa=} $\{\aut\in\NFA:\lang{\aut}=\als^* \}$: 
Deciding whether a given NFA is universal is a PSPACE-complete problem, \cite{HoMoUl:2001}.
%\item{\uunfa:}
%Deciding whether a given unary NFA is universal is an NP-complete problem, \cite{FerKre:2017}.
\item{\ubnfa=} $\{\autb\in\BNFA:\lang{\autb}=\als^\ell, \text{ where $\ell$ is the word length of \autb} \}$: 
Deciding whether a given block NFA  of some word length $\ell$ accepts all words of  length $\ell$ is a coNP-complete problem, \cite{KMR:2018}.
\item{\umaxlennfa=} $\{(\aut,\ell):\aut\in\NFA,\ell \text{ is unary in } \N,\, \lang{\al_s^{\le\ell}}\subseteq\lang{\aut}\}$: Deciding whether $\lang{\al_s^{\le\ell}}\subseteq\lang{\aut}$, for given $\aut\in\NFA$ and \emshort{unary} $\ell\in\N_0$, is coNP-complete, \cite{FerKre:2017}.
\item{\adfasubsetnfa=} $\{(\aut,\autb):\aut\in\NFA,\autb\in\ADFA,\,\lang{\autb}\subseteq\lang{\aut} \}$: Deciding whether $\lang{\autb}\subseteq\lang{\aut}$, for given $\aut\in\NFA$ and $\autb\in \ADFA$ is PSPACE-complete---see below Remark~\ref{REM:adfa}. 
\item{\edfai=} $\{(\aut_1,\ldots,\aut_n): n\in\N, \aut_i\in\DFA,\,\cap_{i=1}^n\lang{\aut_i}=\emptyset\}$: Deciding whether the intersection of given DFAs is empty is PSPACE-complete, \cite{FerKre:2017}. Note that the problem remains hard even if we know that the languages of the given DFAs belong to low levels of the dot-depth or the Straubing-Th\'erien hierarchies  \cite{AFHHJOW:2021}.
%\item{\ebdfai=} $\{(\autb_1,\ldots,\autb_n): n\in\N, \autb_i\in\BDFA,\,\cap_{i=1}^n\lang{\autb_i}=\emptyset\}$: Deciding whether the intersection of given block DFAs is empty is coNP-complete.
\end{description}

\begin{remark}\label{REM:adfa}
	The problem \adfasubsetnfa is PSPACE-hard. This follows when we see that \unfa can be reduced to it using the fact that $\lang{\aut}=\als^*$ iff $\als^*\subseteq\lang{\aut}$. The problem is in PSPACE: as $\autb$ is acyclic, one can enumerate all words of $\lang{\autb}$, \cite{AckSha:2009}, testing whether each one is in $\lang{\aut}$; this process works within polynomial space.
\end{remark}

\pssn 
\textbf{Probability distributions.}
	Let $X$ be a countable set. A \emdef{probability distribution} on $X$ is a function $\dd:X\to[0,1]$ such that 
\begin{equation}\label{eq:distrib}\sum_{x\in X}\dd(x)=1.\end{equation}
The \emdef{domain} of \dd, denoted by $\dom\dd$, is the subset $\{x\in X\mid\dd(x)>0\}$ of $X$. 
%If the set $X$ is finite then the probability distribution is called \emdef{finite}. 
If  $X=\{x_1,\ldots,x_\ell\}$, for some $\ell\in\N$, then we write 
$$D=\big(D(x_1),\ldots,D(x_\ell)\big).$$

Following \cite{Gol:1992},  we have the following definition.

\begin{definition}\label{def:prob}
    Let $\dd$ be a probability distribution on $X$. For any subset $S$ of $X$, we define the quantity
\begin{equation}\label{eq:uindex}\dd(S)=\sum_{x\in S}\dd(x)\end{equation}
and refer to it as \emdef{the probability that a randomly selected element from \dd is in $S$.} The following notation, borrowed from cryptography, means that $x$ \emdef{is randomly selected from} $\dd$:
\[
x\select\dd.
\]
\end{definition}
\begin{remark}\label{lem:wd} 
	Let \dd be a probability distribution on some countable set $X$. The next statements follow from \eqref{eq:distrib} and \eqref{eq:uindex}.
\begin{enumerate}
  \item $\dd(\dom\dd)=1$.
  \item For any  subsets $K$ and $L$ of $X$, if $K\cap L=\es$ then $\dd(K\cup L)=\dd(K)+\dd(L)$.
  \item For any subsets $K$ and $L$ of $X$, if $K\sse L$ then $\dd(K)\le\dd(L)$.
%  \item For any language $L$, $\wrd(L)=\wrd(L\cap\dom\wrd)$.
\end{enumerate}
\end{remark}
The author of \cite{Gol:1992} considers three families of probability distributions on $\N_0$ that are meaningful in information and/or number theory. These distribution families are called uniform, \emdef{Lambert} and \emdef{Dirichlet}, and are defined, respectively, as follows, where $d\in\Ns0,M\in\N, z\in(0,1)$ and $t\in(1,+\infty)$ are related parameters. 
\begin{description}
\item{\emph{Uniform:}} $\ldun M(n)=1/M$ for $n< M$, and $\ldun M(n)=0$ otherwise.
\item{\emph{Lambert:}} $\ldla{1/z,d}(n)=(1-z)z^{n-d}$ for $n\ge d$, and $\ldla{1/z,d}(n)=0$ otherwise.
\item{\emph{Dirichlet:}} $\lddi{t,d}(n) = (1/\zeta(t))(n+1-d)^{-t}$ for $n\ge d$, where $\zeta$ is the Riemann zeta function, and $\lddi{t,d}(n)=0$ otherwise.
\end{description}
In fact \cite{Gol:1992} considers distributions on $\N$, but here we use $\N_0$ instead as we intend to apply these distributions to modelling lengths of words, including possibly the empty word \ew whose length is 0. We also note that \cite{Gol:1992} considers $\ldla{1/z,d}$ and $\lddi{t,d}$ only for the case where the displacement $d=1$. We also note that in \cite{Gol:1970} the same author considers the Dirichlet distribution to be the basis  where  \emlong{``many heuristic probability arguments based on the fictitious uniform distribution on the positive integers become rigorous statements.''}
\begin{definition}\label{def:ldistr}\if\DRAFT1\NOTE{def:ldistr}\fi
	We shall call any probability distribution \ld on $\N_0$ a \emdef{length distribution}. Then, as all values $\ld(n)$ are numeric, \ld can be viewed as a random variable, and the \emdef{expected value} of a length distribution \ld is well-defined and denoted by \ev{\ld}.
\end{definition}
If $t>2$, the expected value of $\lddi{t,1}$ is finite and equal to $\zeta(t-1)/\zeta(t)$, \cite{Gol:1970}. Using standard tools in series manipulation and the fact that $\sum_{i\in\N}iz^i=z/(1-z)^2$, we have the following lemma.
\begin{lemma}\label{lem:exp:length}
	Let $d\in\Ns0,z\in(0,1)$ and $t\in(2,+\infty)$. We have that 
	\[\ev{\ldla{1/z,d}}=d+\frac{1}{1/z-1} \qtim{and} \ev{\lddi{t,d}}=d+\frac{\zeta(t-1)}{\zeta(t)}-1.\]
\end{lemma}

%%%%%%%%%%%%%%%%%%%%%%%%%%%%%%%%%%%%%%%%%%%%%%%%%%%%%%%%%
\section{Randomized Approximation of [0,1]-value problems}\label{sec:randappr}
We consider problems for which every instance\footnote{Following the presentation style of \cite[pg 193]{Gold:2008}, we refrain from cluttering the notation with the use of a variable for the set of instances.} $x$ has a value $\val(x)\in[0,1]$ and we are interested in those instances $x$ for which $\val(x)=1$. Our main set of instances is the set of NFAs (or subsets of that) and the main value function $\val$ is the universality index of NFAs, which is defined in Section~\ref{sec:univ}. However for the purposes of this section, our sample set of instances is $\BNFA[2]$ = all block NFAs over the alphabet $\{0,1\}$, and the $[0,1]$-valued function $v$ is such that $v(\aut)=|\lang\aut|/2^n$, where $n$ is the  word length of the block NFA \aut. In general, for a fixed but arbitrary \emlong{$[0,1]$-valued  function $\val$}, we define the language (problem)
\[
L_{\val}=\{x:\val(x)=1\}.
\]
Deciding whether a given instance $x$ is in $L_{\val}$ might be hard, but \emlong{we assume that we are happy if we know whether  $\val(x)\ge1-\err$, for some appropriate \emdef{tolerance} $\err\in\zerone$}. So we define the following \emshort{approximation} language  for $L_{\val}$:
\[
L_{\val,\err}=\{x:\val(x)\ge1-\err\}.
\]

\begin{remark}\label{R:approx}
One can verify that $L_{\val}=\bigcap_{\err\in(0,1)}L_{\val,\err}$; hence $L_{\val}$ can be approximated as close as desired via the languages $L_{\val,\err}$.
\end{remark}

Unfortunately deciding $L_{\val,\err}$ can be harder than deciding $L_{\val}$, as shown in the proof of the next theorem---the proof  can be found further below.

\begin{theorem}\label{TH:fixed:hard}
	The following problem about block NFAs is coNP-hard  
	\[
	B_{\delta}=\{\aut\in\BNFA[2]: \frac{|\lang\aut|}{2^n} \ge \delta, \text{ where $n$ = word length of $\aut$}\},
	\] 
	for any (fixed) $\delta\in(0,1)$ that is computable within polynomial   time\footnote{A real $x\in(0,1)$ is computable if there is an algorithm that takes as input a positive integer $n$ and computes the  $n$-th bit of $x$. It is computable within polynomial time if the algorithm works in  time $O(n^k)$, for some fixed $k\in\N_0$, when the input $n$ is given in unary.}. 
\end{theorem}

\pnsn
Another idea then is to show that $L_{\val}$ is in the class co\RP, that is, there is a  polynomial \emshort{randomized} algorithm $A(x)$ such that 
\pssi if $x\in L_{\val}$ then $A(x)=\true$  (with probability 1), and 
\pssi if $x\notin L_{\val}$ then $A(x)=\false$ with probability\footnote{Many authors specify this probability to be at least 2/3, but  they state that any value $\ge1/2$ works \cite{Gold:2008,ArBa:2009}.} at least 3/4. 
\pssn
However, as $L_{\val}$ can be hard, it is unlikely that it is in the class co\RP.

\pmsn
The next idea is to devise an approximating algorithm for $L_{\val}$ via $L_{\val,\err}$. As stated in \cite[pg 417]{Gold:2008}, \emlong{``The answer to [what constitutes a "good" approximation] seems intimately related to the specific computational task at hand...the importance of certain approximation problems is much more subjective...[which] seems to stand in the way of attempts at providing a comprehensive theory of natural approximation problems.''} It seems that the following approximation method is meaningful. Although our domain of interest involves NFAs, the below definition is given for any set of instances and refers to a fixed but arbitrary [0,1]-valued function $v$ on these instances.

\begin{definition}\label{D:PRAA}
	Let ${\val}$ be [0,1]-valued function. A \emdef{polynomial  approximation (PAX)} algorithm for $L_{\val}$ is an algorithm  $A(x,\err)$ such that 
	\begin{itemize}
		\item if $x\in L_{\val}$ then $A(x,\err)=\true$;
		\item if $x\notin L_{\val,\err}$ then $A(x,\err)=\false$;
		\item $A(x,\err)$ works within  polynomial time {w.r.t.} $1/\err$ and the size of $x$.
	\end{itemize}
\end{definition}

\pnsn\textbf{Explanation.}
	In the above definition, if $A(x,\err)$ returns \false then $x\notin L_{\val}$, that is, $\val(x)<1$ . If $A(x,\err)$ returns \true then  $x\in L_{\val,\err}$, that is, $\val(x)\ge 1-\err$. Thus, whenever the algorithm returns the answer \false, this answer is correct and exact; when the algorithm returns \true, the answer is \emlong{correct within the tolerance $\err$}.

\pssi
It turns out that, in general, there are problems for which no approximation  algorithm can do better than the exact algorithms.

\begin{proposition}\label{P:nonapprox}
	There is no polynomial approximation algorithm for the problem \ubnfa, unless P=coNP.
\end{proposition}
\begin{proof}
    It is sufficient to consider the subset of the problem for  BNFAs over the binary alphabet. Given $\aut\in\BNFA[2]$, the question of the problem is equivalent to whether $|\lang\aut|/2^n=1$, where $n$ = the word length of \aut.
	If there were a PAX $A(\aut,\err)$ for this problem then we would decide the problem in polynomial time as follows: find out the word length $n$ of the given BNFA \aut, compute $\err=(1+2^n)^{-1}$ and run $A(\aut,\err)$ to get the desired answer.
\end{proof}

\begin{corollary}
	There is no polynomial approximation algorithm for the problem \umaxlennfa, unless P=coNP.
\end{corollary}

\begin{remark}\label{REM:hard}
	Theorem~\ref{TH:fixed:hard}	 implies that, unless P=coNP, block NFA universality over the binary alphabet cannot be  approximated by some  sequence $\big(B_{\delta_n}\big)$, with $\lim\delta_n=0$ and each  $\delta_n$ being polynomially computable. Based on this observation and on Proposition~\ref{P:nonapprox}, we conclude that, \emlong{in general, it is necessary to add  a randomized aspect to our approximation methods.}  We do this immediately below. We also note that there are in fact cases where a PAX algorithm for a hard problem exists---see Section~\ref{sec:PAXforUniv}. 
\end{remark}

\pnsn
The following definition is inspired from the ``approximate'' algorithmic solution of \cite{KMR:2018} for the task of generating an error-detecting code of $N$ codewords, for given $N$, if possible, or an error-detecting code of less than $N$ codewords which is ``close to'' maximal. 
%Although our domain of interest involves NFAs, the below definition is given for any set of instances and refers to a fixed (but arbitrary) [0,1]-valued function $v$ on these instances.

\begin{definition}\label{D:PRAA}
	Let ${\val}$ be [0,1]-valued function. A \emdef{polynomial randomized approximation (PRAX)} algorithm for $L_{\val}$ is a   randomized algorithm  $A(x,\err)$ such that 
	\begin{itemize}
		\item if $x\in L_{\val}$ then $A(x,\err)=\true$;
		\item if $x\notin L_{\val,\err}$ then $\prob{A(x,\err)=\false}\ge 3/4$;
		\item $A(x,\err)$ works within  polynomial time {w.r.t.} $1/\err$ and the size of $x$.
	\end{itemize}
\end{definition}

\pnsn\textbf{Explanation.}
	In the above definition, if $A(x,\err)$ returns \false then $x\notin L_{\val}$. If $A(x,\err)$ returns \true then probably $x\in L_{\val,\err}$, in the sense that $x\notin L_{\val,\err}$ would imply $\prob{A(x,\err)=\false}\ge 3/4$. Thus, whenever the algorithm returns the answer \false, this answer is correct ($x\notin L_{\val}$); when the algorithm returns \true, the answer is \emlong{correct within the tolerance $\err$} ($x\in L_{\val,\err}$) with  probability $\ge 3/4$. The algorithm returns the wrong answer exactly when it returns \true and $x\notin L_{\val,\err}$, but this happens with probability $<1/4$.

\pmsn\textbf{Use of a PRAX algorithm.}
	The algorithm can be used as follows to determine the approximate membership of  a given $x$ in $L_{\val}$ with a probability that can be as high as desired: Run $A(x,\err)$ $k$ times, for some desired $k$, or until the output is \false. If the output is \true for all $k$ times then $\prob{A(x,\err)=\true \text{ for $k$ times}\mid x\notin L_{\val,\err}}<1/4^k$, that is, the probability of incorrect answer is $<1/4^k$.

\begin{proof} (Of Theorem~\ref{TH:fixed:hard}.)
	We reduce   to $B_{\delta}$ the following known coNP-hard problem
	$\ubnfa[2] =\{\autb\in\BNFA[2]: |\lang\autb|=2^\ell, \text{ where $\ell$ = word length of \autb}\}$.
	We need a  reduction that takes any instance \autb in $\BNFA[2]$, of some word length $\ell\in\N$, and constructs (in polynomial time) an instance \aut in $\BNFA[2]$, of some word length $n\in\N$, such that
	\begin{equation}\label{EQ:reduction}	
	|\lang{\autb}|=2^\ell\quad\text{iff}\quad|\lang{\aut}|\ge2^n\delta.	
	\end{equation}
	The main idea is to make a block NFA $\aut$ that accepts a language $F\cdot\lang{\autb}$ of $|F|\cdot|\lang{\autb}|$ words of length $k+\ell$, where $k$ and $|F|$ depend on $\delta$ and $\ell$. If $\delta$ is of the form $m/2^k$ for some $m,k\in\N$, then $F$ is any language of $m$ words of length $k$, and \eqref{EQ:reduction} holds.  The reduction for the general case of $\delta\not=m/2^k$ is described next, where we use the notation (i) $b_p\triangleq$ the bit at position $p$ in the binary representation of $\delta$, for $p\in\N$; (ii) $m_p\triangleq b_12^{p-1}+b_22^{p-2}+\cdots+b_p$; that is, $m_p$ is the numerator of the fraction $m_p/2^p\in(0,1)$ that results when we cut from $\delta$ all bits after position $p$.
	\begin{enumerate}
		\item Let $k=p_1+\ell$, where $p_1=\min\{p\in\N:b_p=1\}$.
		\item Let $m_k=b_12^{k-1}+b_22^{k-2}+\cdots+b_k$. Note that $2^\ell\le m_k<2^k$.
		\item Let $\autb[m_k]$ be any block NFA accepting a language $F$ of exactly $1+m_k$ words of length $k$, such that $\autb[m_k]$ has exactly one final state $f$. 
		\item Let \aut be the block NFA that results by `concatenating' $\autb[m_k]$ and \autb: change all transitions of $\autb[m_k]$ that go to $f$  to go to the start state of \autb. Note that \aut accepts the language $F\cdot\lang{\autb}$ consisting  of $(1+m_k)|\lang{\autb}|$ words of length $n=k+\ell$.
	\end{enumerate}
	We need to show that \eqref{EQ:reduction} holds and that the above reduction (steps 1--4) can de done within polynomial time with respect to $|\autb|$. That \eqref{EQ:reduction} holds follows from the below observations.
	\begin{itemize}
		\item For any bit position $p$ of $\delta$, we have $m_p/2^p<\delta<(1+m_p)/2^p$. 
		\item The above implies that, for any $p$, there is $x_p\in(0,1)$ such that $\delta=(1+m_p-x_p)/2^p$.
		\item If $|\lang{\autb}|=2^\ell$ then $|\lang{\aut}|=(1+m_k)|\lang{\autb}|>2^k\delta2^\ell=2^n\delta$.
		\item If $|\lang{\aut}|\ge2^n\delta$ then $(1+m_k)|\lang{\autb}|\ge2^k2^\ell(1+m_k-x_k)/2^k$ and then
		\[
		|\lang{\autb}| \ge \big(1-\frac{x_k}{1+m_k} \big)2^\ell\quad \Rightarrow \quad
		|\lang{\autb}| > 2^\ell-1,
		\]
		where the above follows when we recall that $2^\ell\le m_k$. 
	\end{itemize}
	That the above reduction (steps 1--4) is polynomial w.r.t. $|\autb|$ follows when we note that 
	(i) $p_1$ is a constant and $k=O(\ell)$. (ii) $\ell$ is essentially presented in unary as the length of any accepting path of \autb, so  $\ell<|\autb|$; then $\ell$ is stored in binary in a variable that can be used to perform arithmetic operations within polynomial time in steps 1--2. (iv) Step~4 can be done in time $O(|\autb[m_k]|+|\autb|)$. (v) Step~3 can be done in time $O(k^2)$ resulting in $\autb[m_k]$ of size $O(k^2)$ as follows:
	\begin{itemize}
		\item Let $m_k=2^{c_1}+\cdots+2^{c_t}$, where the $c_i$'s are the nonzero bit positions in the binary representation of $m_k$, and such that $t\le k$ and $c_1<\cdots c_t<k$. 
		\item For each $i$, make a `straight line' block NFA $\autb_i$ of $k+1$ states accepting all binary strings $\al_2^{c_i}1^{k-c_i}$.
		\item Make the required block NFA $\autb[m_k]$ to be the `union' of all $\autb_i$'s using a single start  state $s$, a single final state $f$, and connecting $s$ to the  second states of the $\autb_i$'s and connecting the second-last states of the $\autb_i$'s to $f$.
	\end{itemize}
\end{proof}

%%%%%%%%%%%%%%%%%%%%%%%%%%%%%%%%%%%%%%%%%%%%%%%%%%%%%%%%%
\section{Word Distributions}\label{sec:worddistr}
A \emdef{word distribution} \wrd is a probability distribution on $\als^*$, that is,  $\wrd:\als^*\to[0,1]$ such that $\sum_{w\in\als^*}\wrd(w)=1.$ If \aut is an NFA then we use the convention that
\[
\mbox{$\wrd(\aut)$\; means\; $\wrd(\lang{\aut})$}.
\]
The \emdef{domain} and \emdef{length} of \wrd are defined, respectively, as follows:
\[\dom\wrd=\{w\in\als^*\mid \wrd(w)>0\}, \quad 
  \dlen\wrd=\{|w|\mid w\in\dom\wrd\}.\]
  We view $\dlen\wrd$ as a random variable such that $\prob{\dlen\wrd=n}=\prob{w\in\als^n}=\wrd(\als^n)$.
The \emdef{expected length} of \wrd is the quantity
\[
\ev{\dlen\wrd} = \sum_{w\in\als^*}\wrd(w)|w|,
\]
which could be finite or $+\infty$.

\begin{example}\label{ex:unifdistr} 
For a finite language $F$, we write $\uwd{F}$ to denote the \emdef{uniform} word distribution on $F$, that is, $\uwd{F}(w)=1/|F|$ for $w\in F$, and $\uwd{F}(w)=0$ for $w\notin F$. Some important examples of uniform word distributions are: 
\begin{itemize}
\item $\uwd{\als^{\ell}}$, where $\ell$ is any word length. Then, $\uwd{\als^\ell}(w)=1/s^\ell$. 
\item $\uwd{\als^{\le\ell}}$, where $\ell$ is any word length. Then, $\uwd{\als^{\le\ell}}(w)=1/t$, where $t=1+s+\cdots+s^\ell$.
\item $\uwd{\lang{\aut}}$, where $\aut$ is an acyclic NFA. We also simply write $\uwd{\aut}$ for $\uwd{\lang{\aut}}$.
\end{itemize}  
\end{example}

\begin{definition}\label{def:lengthdistr} 
	Let \ld be a length distribution. Then \lbdu{\ld} is the word distribution such that 
	\[\lbdu{\ld}(w) \>=\> \ld(\len w)s^{-\len{w}}.\]
    Any such word distribution is called a \emdef{length-based distribution}.
\end{definition}
\begin{remark}\label{rem:explength}
One can verify that, for any length distribution \ld, the following statements hold true, where $n\in\Ns0$.
\begin{enumerate}
\item $\lbdu{\ld}(\als^n)=\ld(n).$
\item $\lbdu{\ld}(\als^{>n})=\ld\big(\N^{>n}\big).$
\item $\ev{\dlen{\lbdu{\ld}}}=\ev{\ld}.$
%\item $\lbdu{\ld}(L)=\sum_{n\in\N_0}\ld(n)\dens L(n)$, where $\dens L(n)=\frac{|L\cap\als^n|}{|\als^n|}$ is the density of $L$ \cite{SYZS:1992}.
\end{enumerate}
\end{remark}
\begin{example}\label{ex:geomdistr}
	Using the Lambert length distribution $\ldla{\asz,d}(n)=(1-1/s)(1/s)^{n-d}$, we define the Lambert, or \emdef{geometric}, word distribution \wdgd on $\als^*$ such that $\wdgd(w)=0$ if $\len w<d$ and, for $\len w\ge d$,
	\[\wdgd(w) = (1-1/\asz)(1/\asz)^{2|w|-d} .\] 
	Then, for all $n,d\in\N_0$ with $n\ge d$, we have 
	\[ \wdgd(\als^n)=(1-1/s)(1/s)^{n-d},\>\> \wdgd(\als^{>n})=(1/s)^{n+1-d},\>\> \ev{\dlen{\wdgd}}=d+1/(s-1).\]
	In particular, for the  alphabet $\al_2=\{0,1\}$, we have that $\wdgt(\al_2)=1/2$, $\wdgt(\al_2^2)=1/2^2$, etc.
\end{example}
\begin{example}\label{ex:riemdistr}
	Let $t\in(2,+\infty)$. Using the Dirichlet length distribution $\lddi{t,d}(n)=(1/\zeta(t))(n+1-d)^{-t}$, we define the \emdef{Dirichlet} word distribution \wdrd on $\als^*$ such that $\wdrd(w)=0$ if $\len w<d$ and, for $\len w\ge d$,
	\[\wdrd(w) = (1/\zeta(t))(\len w+1-d)^{-t}s^{-\len w}.\] 
	Then, for all $n,d\in\N_0$ with $n\ge d$, we have $\wdrd(\als^n)=(1/\zeta(t))(n+1-d)^{-t},$
	\[ \wdrd(\als^{>n})=1-(1/\zeta(t))\sum_{i=1}^{n+1-d}i^{-t},\quad \ev{\dlen{\wdrd}}=d+\zeta(t-1)/\zeta(t)-1.\]
	In particular, for $t=3$, $d=1$ and  alphabet $\al_2=\{0,1\}$, we have that $\wdrt(\al_2^n)=(1/\zeta(3))n^{-3}$.
\end{example}

\pnsn\textbf{Selecting a word from a distribution.}
We are interested in word distributions \wrd for which there is an efficient (randomized) algorithm %$\selectfrom_\wrd()$ 
that returns  a randomly selected element from \wrd. 
%Moreover, we want \wrd to be polynomial, meaning that, there is $k\in\N_0$ such that $\selectfrom_\wrd()$ returns a word $w$ in time $O(|w|^k)$. 
We shall assume available  (randomized) algorithms as follows.
\begin{itemize}
	\item $\tosscoin(p)$: returns 0 or 1, with probability $p$ or $1-p$, respectively, where $p\in[0,1]$, and the algorithm  works in constant time for most practical purposes---this  is a reasonable assumption according to \cite[pg 134]{ArBa:2009}. 
	\item $\uselect(s,\ell)$: returns a uniformly selected word from $\als^\ell$, and the algorithm works in time $O(\ell)$.
	%\item $\selectwordfin(N)$, where $N=\big(N(\ell_1),\ldots,N(\ell_n)\big)$ is a finite probability distribution on some length set $\{\ell_1,\ldots,\ell_n\}$: returns a randomly selected word from $\lbdu{N}$, that is, a  word $w$ is selected with probability $\lbdu{N}(w)=N(|w|)s^{-|w|}$. This task can be accomplished as follows (i) use $\selectfin(N)$ of Lemma~\ref{lem:selectfin} below to return a word length $\ell_i$; and (ii) use $\uselect(s,\ell_i)$ to return a uniformly selected word from $\als^{\ell_i}$. 
\end{itemize}

\begin{remark}
As in \cite[pg 126]{ArBa:2009}, we assume that basic arithmetic operations are performed in constant time. Even if we relax this assumption and we account for a parameter $q$ for arithmetic precision, the arithmetic operations would require a  polynomial factor in $q$. 
\end{remark}

The next lemma seems to be folklore, but we include it here for the sake of clarity and self-containment.

\begin{lemma}\label{lem:selectfin}
There is a polynomial randomized algorithm $\selectfin(D)$, where 
	$D$ is a finite probability distribution $\big(D(x_1),\ldots,D(x_n)\big)$  on some set $\{x_1,\ldots,x_n\}$, that returns a randomly selected $x_i$ with probability $D(x_i)$. In fact the algorithm works in time $O(n)$ using the assumption of constant cost of $\tosscoin$ and of arithmetic operations. 
\end{lemma}
\begin{proof}
The algorithm works as follows: 
perform  up to $n-1$ coin tosses such that 
\begin{itemize}
\item in the $i$-th coin toss, the  outcome 0 means to return the element $x_i$ and terminate, and the  outcome 1 means to continue to the next coin toss (or return $x_n$ if $i=n-1$); 
\item each coin toss $i$ uses the algorithm $\tosscoin(p_i)$, where $p_1=D(x_1)$ and $p_{i+1}=D(x_{i+1})/\big((1-p_1)\cdots(1-p_i)\big)$.
\end{itemize}
We have that $p_{i}$ = the probability that coin toss $i$ is 0 given that all previous tosses (when $i>1$) are all 1. The outcome $O$ of the algorithm is such that $\prob{O=x_1}=p_1=D(x_1)$ and $\prob{O=x_{i+1}}=p_{i+1}\cdot(1-p_1)\cdots(1-p_i)$.
\end{proof}

\pmsn\textbf{Augmented word distributions.}
Selecting a word from a distribution \wrd with infinite domain $\dom\wrd$ could return a very long word, which can be intractable. For this reason we would like to define distributions on $\als^*\cup\{\none\}$, \emlong{where `\none' is a symbol outside of \als}, which could select the outcome `\none' (no word). These could be versions of   word distributions in which there is a bound on the length of words they can select. 
\begin{definition}\label{def:distr:maxlen}
	An \emdef{augmented} word distribution is a probability distribution on $\als^*\cup\{\none\}$. Let \wrd be a word distribution and let $M\in\N_0$. We define the augmented distribution $\wrdm$ such that
\[
\wrdm(w)=\wrd(w), \mbox{ if $|w|\le M$};\quad 
\wrdm(\none)=\wrd(\als^{>M}).
\]
\end{definition}

%Note that an augmented word distribution $X$ for which $X(\none)=0$ is essentially a word distribution.

\begin{remark}
The probability that $\wrdm$ selects a word longer than $M$ is zero. We have that $\dom\wrdm=(\dom\wrd\cap\als^{\le M})\cup\{\none\}$. Moreover, the following facts about $\wrdm$ and any language $L$ are immediate
\begin{equation}\label{eq:wrdm}
	\wrd(L\cap\als^{\le M})=\wrdm(L\cap\als^{\le M}), \quad
	\wrd(\als^{>M})=\wrdm(\none).
\end{equation}
\end{remark}

\begin{remark}
The proof of Lemma~\ref{lem:selectfin} uses a general formula for computing the quantities $p_i$. However, these quantities can be computed in a much simpler way for specific distributions. For the augmented Lambert distribution $\ldla{s,d}^M$, for instance,  we have that each $p_i=1-1/s$. 
\end{remark}

%%%%%%%%%%%%%%%%%%%%%%%%%%%%%%%%%%%%%%%%%%%%%%%%%%%%%%%%%
\section{Universality Index of NFAs}\label{sec:univ}
Here we intend to define mathematically the informal concept of an ``approximately universal NFA'' with respect to a certain fixed language \ulm. Our motivation comes from coding theory where the codes of interest are subsets of \ulm, and it is desirable that a code is a maximal subset of \ulm. Two typical cases are (i) $\ulm =\als^*$, when variable-length codes are considered, such as prefix or suffix codes; and (ii) $\ulm=\als^n$ for some $n\in\N$, when error control codes are considered. Testing whether a regular code $C$ is a maximal subset of \ulm is a hard problem and, in fact, this problem normally reduces to whether a certain NFA that depends on $C$ accepts  \ulm---see e.g., \cite{DudKon:2012,KMR:2018}. In practice, however, it could be acceptable that a code is ``close'' to being maximal, or an NFA is ``close'' to being universal. 
\pnsi\emlong{Our approach here assumes that the maximum language \ulm is equal to $\dom\wrd$, where \wrd is the word distribution of interest.}

%For any $p\in[0,100]$ we shall write $\upp$ to mean the value $\upar/100$, which is in $[0,1]$. Following ideas from \cite{KMR:2016} about \upp-maximal error-detecting subsets of $\als^n$, we can say that an NFA $\aut$ accepting a subset of $\als^n$ is \upp-universal (with respect to $\als^n$) if $|\lang{\aut}|/|\als^n|\ge\upp$. Next we use probability distributions on words to extend \upp-universality to any NFA.
%
%
\begin{definition}\label{def:uindex}
	Let \wrd be a word distribution, %with $\dom\wrd=\ulm$, %let $L$ be a language, 
	let \aut be an NFA, and let $\upar\in[0,1]$. 
	\begin{itemize}
      \item We say that \aut is \emdef{universal relative} to \wrd, if $\lang\aut=\dom\wrd$. 
      %We say that \aut is block universal if it is a block NFA and $\lang{\aut}=\als^n$, where $n$ is the word length of \aut.
      \item We say that \aut is \emdef{\upar-universal relative} to \wrd, if $\wrd(\aut)\ge\upar$. We call the quantity $\wrd(\aut)$ the \emdef{universality index} of \aut (relative to $\wrd$).
    \end{itemize}
\end{definition}

\begin{example}
	Let $\autb$ be a block NFA. If $\card{\lang\autb}/\asz^\ell\ge\upar$, where $\ell$ is the word length of \autb, then $\autb$ is $p$-universal relative to the uniform distribution on $\als^\ell$  and the quantity $\card{\lang\autb}/\asz^\ell$ is the universality index of \autb.
\end{example}

\begin{remark}\label{rem:uindex}
	The universality index  $\wrd(\aut)$ represents the probability that a randomly selected word from \wrd is accepted by \aut---see Definition~\ref{def:prob}.  When $\wrd(\aut)$ is close to 1 then $\aut$ is close to being universal, that is, $\lang{\aut}$ is close to $\dom\wrd$. The concept of a $\upar$-universal NFA formalizes the loose concept of an approximately universal NFA---see also the next lemma. Thus, for example, we can talk about a 98\%-universal block NFA with respect to the uniform  distribution on $\als^\ell$, where $\ell$ is the word length of the NFA. 
\end{remark}

\begin{remark}\label{rem:muoperator}
The method of \cite{KonMas:2017}  embeds a given $\tr$-code\footnote{Depending on \tr, which is a transducer, one can have prefix codes, suffix codes, infix codes, error control codes.} $K$ into a maximal one by successive applications of a language operator $\mu_{\tr}$ on $K$ which yields supersets $K_i$ of $K$ until these converge to a maximal $\tr$-code. The operation $\mu_{\tr}$ on each $K_i$ (represented as an NFA) can be  expensive to compute and one can simply stop at a step where the current superset $K_i$ is close to maximal, or  according to the concepts of this paper, when the NFA for $\big(\tr(K_i)\cup\tr^{-1}(K_i)\cup K_i\big)$ is close to universal.
\end{remark}
\begin{lemma}\label{lem:uindex}
	If $L$ is universal relative to \wrd, that is $L=\dom\wrd$, then $\wrd(L)=1$, for any word distribution \wrd. Conversely, if there is a word distribution \wrd such that $\wrd(L)=1$, then $L$ is universal relative to $\dom\wrd$.
\end{lemma}
\begin{proof}
	Immediate. 
\end{proof}
%
%
%\begin{remark}\label{rem:uindex2}
Consider the case where $\autb$ is a block NFA of length $\ell$ and $\wrd$ is the uniform word distribution on $\als^\ell$. In this work, we view estimating the universality index of $\autb$ as a \emlong{parameter estimation problem} for finite populations \cite[pg 72]{MiUp:2017}: let \upar be an unknown population parameter (ratio of elements having some attribute over the cardinality of the population). Select $n$ elements from the population (here, $n$ words from $\als^\ell$) and compute $c$, the number of these elements having the attribute of interest (here, words that are in $\lang{\autb}$). Then, $c/n$ is an estimate for the population parameter \upar (here, the estimate is for $\wrd(\autb)$) in the sense that the expected value of the random variable $c/n$ is equal to \upar and 
\begin{equation}\label{eq:chernoff}
\prob{\,|c/n-\upar|>\err} \><\> e^{-n\err^2/2}+e^{-n\err^2/3},
\end{equation}
where $\err>0$ is the acceptable estimation error. The above inequality is given in~\cite{MiUp:2005} and follows from Chernoff bounds. Here we extend the idea of parameter estimation to various distributions on languages. Moreover, we use the simpler Chebysev inequality for bounding the error probability, as it gives in practice a smaller bound than the  one in the above inequality. Let X be a random variable and let $a>0$. The Chebyshev inequality is as follows %\cite{MiUp:2005},
\begin{equation*}
	\prob{\,|X-\ev X|\ge a}\le \sigma^2/a^2,
\end{equation*}
where $\sigma^2$ is the variance of  $X$. When $X$ is the binomial random variable with parameters $n$ = `number of trials' and $p$ = `probability of success in one trial', then $\ev X=np$ and $\sigma^2=np(1-p)$. For $p\in[0,1]$, the maximum value of $p(1-p)$ is 1/4; therefore, the above inequality becomes as follows:
\begin{equation}\label{eq:cheb}
	\prob{\,|X-\ev X|\ge a}\le n/(4a^2).
\end{equation}
%\end{remark}
%
%
%\pssn
%\textbf{\upp-Univeral languages and NFAs.} Next we define the concept of  \upp-universality for languages and NFAs.
%\begin{definition}\label{def:puniv}
%	Let $\upar\in[0,100]$. A language $L\sse\dom\wrd$ is called \emdef{\upp-universal}, if  $\wrd(L)\ge\upp$. An NFA \aut is called \emdef{\upp-universal}, if  $\lang\aut$ is \upp-universal.
%\end{definition}
%
%
%\begin{remark}
%From the discussion on probability distributions in Section~\ref{sec:notation}, it follows that when we say that $L$ is $90\%$-universal, for instance, it means that that a randomly selected word from $\dom\wrd$ will be in the language $L$ with probability at least 90\%, where the word is selected according to \wrd.  
%\end{remark}
%
%
%
\begin{figure}[ht]
\begin{center}
\parbox{0.75\textwidth}
{
\hspace*{0.3\algoindent} %Algorithm 
$\uiestim_{\wrd} (\aut, n)$\if\DRAFT1\NOTE{fig:param:est}\fi
\pssn \hspace*{\algoindent}
%$n$ := 1 + $\Big\lfloor\,1/\Big(4\upar (1-\upar)^2\Big)\,\Big\rfloor$;
%$\ell$ := the word length of $\lang\aut$;
%\\ \hspace*{\algoindent}
%A := the alphabet of \aut;
%\\ \hspace*{\algoindent}
%$n$ := $\lceil1/\err^2\rceil$;
%\\ \hspace*{\algoindent}
cnt := 0;
\\ \hspace*{\algoindent}
i := 0;
\\ \hspace*{\algoindent}
while (i $<$ $n$):
\\\hspace*{1.7\algoindent} 
$w$ $\select$ $\wrd$;
\\ \hspace*{1.7\algoindent} 
i := i+1;
\\ \hspace*{1.7\algoindent}
if ($w$ $\in$ $\lang{\aut}$) cnt := cnt+1;
\\ \hspace*{\algoindent} 
return cnt / $n$;
}
\parbox{0.90\textwidth}{\caption{This random process refers to a particular  word distribution  $\wrd$. It returns an estimate of the universality index $W(\aut)$ of the given NFA \aut.}\label{fig:param:est}}
\end{center}
\end{figure}
\begin{lemma}\label{lem:univ:estim}
	Let \aut be an NFA, let $\wrd$ be a word distribution, and let $\upar,\uparg\in[0,1]$ with $\upar>\uparg$. Consider the random process $\uiestim_\wrd(\aut,n)$ in Fig.~\ref{fig:param:est}, and let \cnt be the random variable for the value of \emph{cnt} when the algorithm returns. If $\wrd(\aut)<\uparg$ then
	$
	\prob{\cnt/n\ge \upar} \le \frac{1}{4n(\upar-\uparg)^2}.
	$ 
\end{lemma}
\begin{proof}
	First note that \cnt is binomial: the number of successes (words in $\lang{\aut}$) in $n$ trials. Thus, $\ev{\cnt}=n\wrd(\aut)$. Now assume that $\wrd(\aut)<\uparg$. We have:
\begin{eqnarray*}
& &\prob{\cnt/n\ge \upar} \> = \>  
   \prob{\cnt-n\wrd(\aut)\ge n\upar-n\wrd(\aut)} \\ 
&\le & \prob{\,|\cnt-n\wrd(\aut)|\ge n\upar-n\wrd(\aut)\,} \\ 
&\le & \prob{\,|\cnt-n\wrd(\aut)|\ge n\upar-n\uparg\,} \le
       \frac{1}{4n(\upar-\uparg)^2},
\end{eqnarray*}
where we have used inequality~\eqref{eq:cheb}. 
\end{proof}

In Section~\ref{sec:urNFA} we give a polynomial randomized approximation algorithm (PRAX) for testing universality of block NFAs, which is based on the random process in Fig.~\ref{fig:param:est}. That process, however, cannot lead to a PRAX for the universality of NFAs accepting infinite languages, as the selection $w\select\wrd$ could produce a word of exponential length. In Fig.~\ref{fig:param:est2} we modify that process so that a selected word cannot be longer than a desired $M\in\N_0$---in Section~\ref{sec:uNFA} we investigate how this can lead to a PRAX for the universality of any NFA relative to tractable word distributions.
\begin{figure}[ht]
\begin{center}
\parbox{0.85\textwidth}
{
\hspace*{0.3\algoindent} %Algorithm 
$\uiestimml_{\wrd}(\aut, n, M)$  
\pssn \hspace*{\algoindent}
cnt := 0;
\\ \hspace*{\algoindent}
i := 0;
\\ \hspace*{\algoindent}
while (i $<$ $n$):
\\\hspace*{1.7\algoindent} 
$w$ $\select$ $\wrdm$;
\\ \hspace*{1.7\algoindent} 
i := i+1;
\\ \hspace*{1.7\algoindent}
if \big($w=\none$ \;or\; $w$ $\in$ $\als^{\le M}\cap\lang{\aut}$\big) cnt := cnt+1;
\\ \hspace*{\algoindent} 
return cnt / $n$;
}
\parbox{0.90\textwidth}{\caption{This random process refers to a particular  word distribution  $\wrd$. It returns an estimate of $\wrdm(\none)+\wrdm(\als^{\le M}\cap\lang\aut)$, which is equal to $\wrd(\als^{>M})+\wrd(\als^{\le M}\cap\lang\aut)$---see \eqref{eq:wrdm}. When $M$ is chosen such that $\wrdm(\none)$ is small enough, then the returned quantity cnt/$n$ can be an acceptable estimate of $\wrd(\aut)$.}\label{fig:param:est2}}
\end{center}
\end{figure}

\begin{lemma}\label{lem:univ:maxlen:estim}
	Let \aut be an NFA, let  $\wrd$ be a  word distribution, let $M\in\N_0$, and let $\upar,\uparg\in[0,1]$ such that $\upar>\uparg+\wrd(\als^{>M})$. Consider the random process $\uiestimml_{\wrd}(\aut,n,M)$ in Fig.~\ref{fig:param:est2}, and let \cnt be the random variable whose value is equal to the value of  \emph{cnt} when the algorithm returns. If $\wrd(\aut)<\uparg$ then
	$$
	\prob{\cnt/n\ge \upar} \>\le\> \frac{1}{x}+\frac{1}{4n\big(\upar-\uparg-x\wrd(\als^{>M})\big)^2},\quad
	\mbox{ for all } x\in\big(1,\frac{\upar-\uparg}{\wrd(\als^{>M})}\big).$$
\end{lemma}
\begin{proof}
	Referring to the $n$ selections $w\select\wrdm$ in $\uiestimml_\wrd(\aut,n,M)$, let \icnt be the random variable for the number of selections that are in $\lang\aut\cap\als^{\le M}$, and let \bcnt be the random variable for the number of selections equal to \none. Then, $\cnt=\icnt+\bcnt$. Now note that (i) \icnt is binomial: the number of successes (words in $\lang{\aut}\cap\als^{\le M}$) in $n$ trials, and (ii) \bcnt is binomial: the number of successes (selections \none) in $n$ trials. Thus, using \eqref{eq:wrdm}, we have
	$$\ev{\icnt}=n\wrd\big(\lang\aut\cap\als^{\le M}\big)\le n\wrd(\aut), \quad \ev{\bcnt}=n\wrd(\als^{>M}).
	$$
	Now assume that $\wrd(\aut)<\uparg$, and let $x$ be a number with $1<x<(p-g)/\wrd(\als^{>M})$. We have:
\begin{eqnarray*}
& & \prob{\cnt/n\ge \upar} =
    \prob{\icnt+\bcnt\ge n\upar} = \prob{\icnt\ge n\upar-\bcnt} \\ 
&=& \prob{\icnt\ge n\upar-\bcnt \mbox{ and } \bcnt>xn\wrd(\als^{>M})} 
  +\prob{\icnt\ge n\upar-\bcnt \mbox{ and } \bcnt\le xn\wrd(\als^{>M})}\\
&\le & \prob{\bcnt>xn\wrd(\als^{>M})} + 
       \prob{\icnt\ge n\upar-\bcnt \mbox{ and } n\upar-\bcnt\ge n\upar-xn\wrd(\als^{>M})} \\
&\le & \prob{\bcnt>xn\wrd(\als^{>M})} + 
       \prob{\icnt\ge n\upar-xn\wrd(\als^{>M})} \\
&\le & \ev{\bcnt}/\big(xn\wrd(\als^{>M})\big)  +
       \prob{\icnt-\ev{\icnt}\ge n\upar-\ev{\icnt}-xn\wrd(\als^{>M})} \\          
&\le & 1/x \>+\> \prob{\,|\icnt-\ev{\icnt}|\ge n\upar-n\uparg-xn\wrd(\als^{>M})\,} \\ 
&\le & \frac{1}{x}+\frac{1}{4n\big(\upar-\uparg-x\wrd(\als^{>M})\big)^2},
\end{eqnarray*}
where we have used Markov's inequality ``$\prob{\bcnt>a}<\ev{\bcnt}/a$, for all $a>0$'', as well as inequality~\eqref{eq:cheb}.
\end{proof}
%
%
%\begin{example}\label{ex:univ1} 
%	A language $L$ is called an \emdef{infix code} if no infix of an $L$-word is equal to another $L$-word: $u\in\infix(v)$ implies $u=v$, for any $u,v\in L$. The language $L$ is a \emdef{maximal} infix code, if $L\cup\{w\}$ is not an infix code for all words $w\notin L$. In \cite{DudKon:2012} it is shown that deciding whether a given acyclic NFA \aut accepts a maximal infix code is a coNP-hard problem, and that deciding whether any NFA \aut (not necessarily acyclic) accepts a maximal infix code reduces to testing whether 
%	\[\infix(\lang\aut)\cup\als^*\,\lang\aut\,\als^*=\als^*.\]
%	One can construct an NFA \autb accepting the language $\infix(\lang\aut)\cup\als^*\,\lang\aut\,\als^*$, so the infix maximality of $\lang{\aut}$ reduces to the universality of the NFA \autb. As universality testing is hard, one would at least like to know how close \autb is to being universal. For example, let $L=0^*1$. One can verify that $\infix(L)\cup\als^*\,L\,\als^*=\als^*$. Now let $L=10^*1$. 
%	\pssi\sk{FIX this using suffix code property and an automaton.}
%\end{example}

%%%%%%%%%%%%%%%%%%%%%%%%%%%%%%%%%%%%%%%%%%%%%%%%%%%%%%%%%
\section{Randomized Approximation of NFA problems relative to Uniform Distributions}\label{sec:urNFA}

In this section we consider polynomial randomized approximation algorithms for the problems \adfasubsetnfa, \ubnfa, \umaxlennfa.  As discussed below, the latter two problems are essentially special cases of the problem \adfasubsetnfa, but they can also be answered using a couple of more standard tools leading to more efficient algorithms.

\begin{figure}[ht]
\begin{center}
\parbox{0.43\textwidth}
{
\hspace*{0.3\algoindent} %Algorithm 
$\mathsf{ADFASubsetNFA}(\aut, \autb, \err)$  
\pssn \hspace*{\algoindent}
$n$ := $\lceil1/\err^2\rceil$;
\\ \hspace*{\algoindent}
i := 0;
\\ \hspace*{\algoindent}
while (i $<$ $n$):
\\\hspace*{1.7\algoindent} 
$w$ := $\uselect(\autb)$;
\\ \hspace*{1.7\algoindent} 
i := i+1;
\\ \hspace*{1.7\algoindent}
if ($w\notin\lang{\aut}$) 
return \false;
\\ \hspace*{\algoindent} 
return \true;
\\ \hspace*{\algoindent} 
}
\qquad
\parbox{0.47\textwidth}
{
\hspace*{0.3\algoindent} %Algorithm 
$\mathsf{ADFASubsetNFA}(\aut, \autb, \err)$   
\pssn \hspace*{\algoindent}
$n$ := $\lceil1/\err^2\rceil$;
\\ \hspace*{\algoindent}
i := 0; \quad cnt := 0;
\\ \hspace*{\algoindent}
while (i $<$ $n$):
\\\hspace*{1.7\algoindent} 
$w$ := $\uselect(\autb)$;
\\ \hspace*{1.7\algoindent} 
i := i+1;
\\ \hspace*{1.7\algoindent}
if ($w$ $\in$ $\lang{\aut}$) cnt := cnt + 1;
\\ \hspace*{\algoindent} 
if (cnt $<n$) return \false 
\\ \hspace*{\algoindent} 
else return \true;
}
\parbox{0.90\textwidth}{\caption{On the left is the PRAX algorithm for the problem \adfasubsetnfa: whether the language of the given acyclic DFA $\autb$ is a subset of the language of the given NFA $\aut$. This is equivalent  to whether $\lang{\autb}\subseteq\lang{\aut}\cap\lang{\autb}$. The function $\uselect(\autb)$ returns a uniformly selected word from $\lang{\autb}$. The version on the right is logically equivalent; it mimics the process in Fig.~\ref{fig:param:est} and is intended to give a more clear explanation of correctness.}\label{fig:algo0}}
\end{center}
\end{figure}

\begin{lemma}\label{L:select:bnfa}
	Selecting uniformly at random an accepting word of a given  ADFA \aut can be done in polynomial time.
\end{lemma}
\begin{proof}
    The statement can be shown using results from \cite{BerGim:2012}. However, we give here a simple self-contained presentation. \emshort{First}, let $N(q)$ be the number of words accepted by \aut from the start state $s$ to state $q$. 
    We have that $N(s)=1$ and then, for each state $q$ in breadth-first order, $N(q)$ is the sum of $N(p)$ for all transitions $(p,\sigma,q)$ leading into $q$, where each computed value $N(q)$ is recorded so that it can be reused. Let $F$ be the set of final states of \aut and let $N_F=\sum_{f\in F}N(f)$, which is equal to $|\lang{\aut}|$. 
    \emshort{Then}, selecting a word $w\in\lang{\aut}$ can be done in two steps. The first step is to use $\selectfin$ to select a final state $f$ from the distribution $\big(N(f)/N_F\big)_{f\in F}$. The second step is to select a word $w$ accepted by \aut at the final state $f$. 
    Each symbol $\sigma$ of $w$ is selected starting from the last one, as follows. Let $T_f$ be the set of transitions leading to state $f$. Use again $\selectfin$ to select one transition $(p,\sigma,f)\in T_f$ from the distribution that consists of the values $N(p)/N(f)$ for  all $(p,\sigma,f)\in T_f$. 
    Then, the last symbol of $w$ is $\sigma$. Repeat the same process, for $f\leftarrow p$,  selecting the next symbol of $w$, until the start state $s$ is encountered.
\end{proof}

\begin{theorem}\label{th:finapprox}
	Algorithm $\mathsf{ADFASubsetNFA}(\aut, \autb, \err)$ is a polynomial randomized approximation algorithm for \adfasubsetnfa.
\end{theorem}
\begin{proof}
    First we note that \adfasubsetnfa can be expressed as follows as a [0,1]-value problem
    \[
    \adfasubsetnfa=\Big\{(\aut,\autb):\aut\in\NFA,\autb\in\ADFA,\;v(\aut,\autb)=1
    %\frac{|\lang{\aut}\cap\lang{\autb}|}{|\lang{\autb}|}=1 
    \Big\},
    \]
    where $v(\aut,\autb)=\frac{|\lang{\aut}\cap\lang{\autb}|}{|\lang{\autb}|}$;
    therefore the problem $\adfasubsetnfa_{\err}$ is well-defined.
    For brevity we write  $A(\aut, \autb, \err)$ to refer to $\mathsf{ADFASubsetNFA}(\aut, \autb, \err)$. We consider the three conditions of Definition~\ref{D:PRAA}.
	The \emshort{third} condition about the time complexity follows when we note that
	(i) testing whether a word $w$ is accepted by an NFA \aut can be done in time $O(|w|\sz\aut)$; and 
	(ii)  selecting uniformly at random a word from an acyclic DFA $\autb$ can be done in polynomial time (see Lemma~\ref{L:select:bnfa}). 
	For the \emshort{first} condition of Definition~\ref{D:PRAA}, if $\lang{\autb}\subseteq\lang{\aut}$ then every selected word $w$ is in $\lang\aut$, so the algorithm will return \true. 
	For the \emshort{second} condition, assume that $|\lang{\aut}\cap\lang{\autb}|/|\lang{\autb}|<1-\err$. Consider the version of the algorithm on the right and the random process in Lemma~\ref{lem:univ:estim} and assume that it selects exactly the same words $w$ as $A(\aut,\autb,\err)$ does. 
	Then, algorithm $A(\aut,\autb,\err)$ returns \true if and only if the random variable \cnt in Lemma~\ref{lem:univ:estim} takes the value $n$. Moreover, using $\upar=1$ and $\uparg=1-\err$ in Lemma~\ref{lem:univ:estim}, we have
\begin{eqnarray*}
%& &
\prob{A(\aut,\autb,\err)=\true} =\prob{\cnt=n} &=& \prob{\cnt/n\ge1}\\
%\\ &\le & 
&\le& \frac{1}{4n\big(1-(1-\err)\big)^2} =  \frac{1}{4n\err^2}\>\le 1/4.
\end{eqnarray*}
\end{proof}

The next corollaries follow from the  above theorem; however, using a more self-contained choice of tools we get more efficient algorithms with estimates of their time complexity.

\begin{figure}[ht]
\begin{center}
\parbox{0.45\textwidth}
{
\hspace*{0.3\algoindent} %Algorithm 
$\ubalgo(\aut, \err)$  
\pssn \hspace*{\algoindent}
%$n$ := 1 + $\Big\lfloor\,1/\Big(4\upar (1-\upar)^2\Big)\,\Big\rfloor$;
$\ell$ := the word length of $\lang\aut$;
\\ \hspace*{\algoindent}
$n$ := $\lceil1/\err^2\rceil$;
\\ \hspace*{\algoindent}
i := 0;
\\ \hspace*{\algoindent}
while (i $<$ $n$):
\\\hspace*{1.7\algoindent} 
$w$ := $\uselect(s,\ell)$;
\\ \hspace*{1.7\algoindent} 
i := i+1;
\\ \hspace*{1.7\algoindent}
if ($w$ $\notin$ $\lang{\aut}$) return \false;
\\ \hspace*{\algoindent} 
return \true;
}
\qquad
\parbox{0.45\textwidth}
{
\hspace*{0.3\algoindent} %Algorithm 
$\umaxlenalgo(\aut,\ell, \err)$  \if\DRAFT1\NOTE{fig:algo1}\fi
\pssn \hspace*{\algoindent}
$t$ := $1+s+\cdots+s^\ell$;
\\ \hspace*{\algoindent}
$N$ := $(1/t,s/t,\ldots,s^\ell/t)$;
\\ \hspace*{\algoindent}
$n$ := $\lceil1/\err^2\rceil$;
\\ \hspace*{\algoindent}
i := 0;
\\ \hspace*{\algoindent}
while (i $<$ $n$):
\\ \hspace*{1.7\algoindent} 
$k$ := $\selectfin(N)$;
\\\hspace*{1.7\algoindent} 
$w$ := $\uselect(s,k)$;
\\ \hspace*{1.7\algoindent} 
i := i+1;
\\ \hspace*{1.7\algoindent}
if ($w$ $\notin$ $\lang{\aut}$) return \false;
\\ \hspace*{\algoindent} 
return \true;
}
\parbox{0.90\textwidth}{\caption{$\ubalgo$ decides approximate block NFA universality (see Corollary~\ref{th:ubapprox}) and $\umaxlenalgo$ decides approximate up to a maximum length NFA universality  (see Corollary~\ref{th:umaxlenapprox}).}\label{fig:algo1}}
\end{center}
\end{figure}

\begin{corollary}\label{th:ubapprox}
	Algorithm $\ubalgo(\aut, \err)$ in Fig.~\ref{fig:algo1} is a polynomial randomized approximation algorithm for block NFA universality and works in time $O\big(\ell\,\sz\aut(1/\err)^2\big)$, where $\ell$ is the word length of \aut.
\end{corollary}
\begin{proof}
	The existence of a polynomial randomized approximation algorithm for block NFA universality follows from  the algorithm $\mathsf{ADFASubsetNFA}(\aut, \autb, \err)$ of Theorem~\ref{th:finapprox} when we note that given block NFA \aut of some word length  $\ell$, one can construct in time $O(\ell)$ a block (hence, acyclic) DFA \autb accepting the language $\al_s^{\ell}$. Here however, step $\uselect(\autb)$ of $\mathsf{ADFASubsetNFA}$ can be replaced by the simpler process of selecting uniformly a word of length $\ell$.
\end{proof}

\pnsn\textbf{Use of the algorithm $\ubalgo(\aut, \err)$.}
Suppose that we want to test whether a block NFA \aut of some word length $\ell$ is universal relative to the uniform distribution on $\als^\ell$, and that we allow a 2\% approximation tolerance, that is, we consider it acceptable to say that \aut is universal when it is in fact 98\%-universal. Then we run the algorithm using  \err = 0.02. If \aut is universal, then the algorithm correctly returns \true. If \aut is not 98\%-universal, then the probability that the algorithm returns \true is at most 1/4. Note that for this choice of arguments, the loop would iterate at most 2500 times.

\begin{corollary}\label{th:umaxlenapprox}
	Algorithm $\umaxlenalgo(\aut,\ell, \err)$ in Fig.~\ref{fig:algo1} is a polynomial randomized approximation algorithm for \umaxlennfa. In fact the algorithm works in time $O\big(\ell\,|\aut|(1/\err)^2\big)$ under the assumption of constant cost of $\tosscoin$ and of arithmetic operations.
\end{corollary}
\begin{proof}
	The existence of a polynomial randomized approximation algorithm for \umaxlennfa follows from    the algorithm $\mathsf{ADFASubsetNFA}(\aut, \autb, \err)$ of Theorem~\ref{th:finapprox} when we note that given $\ell$ in unary, one can construct in time $O(\ell)$ an acyclic DFA \autb accepting the language $\al_s^{\le\ell}$. Here however, step $\uselect(\autb)$ of $\mathsf{ADFASubsetNFA}$ can be replaced by the process of selecting uniformly a word length $k\in\{0,1,\ldots,\ell\}$ according to the distribution
	\[\big(|\al_s^0|/t,\;|\al_s^1|/t,\ldots,|\al_s^\ell|/t\big)\]
	and then selecting uniformly a word of length $k$.
\end{proof}

%%%%%%%%%%%%%%%%%%%%%%%%%%%%%%%%%%%%%%%%%%%%%%%%%%%%%%%%%
\section{Randomized Approximation of NFA Universality}\label{sec:uNFA}
Here we present an analogue to the uniform distribution algorithms  for the case where the NFA accepts an infinite language and universality is with respect to some word distribution \lbdu\wdt. The approximation algorithm of this section is based on the random process in Fig.~\ref{fig:param:est2} and requires that the distribution \lbdu\wdt be tractable, which loosely speaking means that  words longer than a certain length $M=M(\err)$ have low probability and can be ignored when one wants to approximate the universality index of the given NFA  \emlong{within a given tolerance \err}---recall, this approach is consistent with our interpretation of languages in the context of coding and information theory.

\begin{definition}\label{def:tractable}
	A length distribution \wdt is called \emdef{tractable}, if the following conditions hold true.
\begin{enumerate}
  \item For all $\err\in(0,1)$, there is $M\in\Ns0$ such that $\wdt(\N^{>M})\le\err$, $M$ is of polynomially bounded magnitude  w.r.t. $1/\err$, that is, $M=O\big((1/\err)^k\big)$ for some $k\in\N_0$, and there is an algorithm $\maxlen_\wdt(\err)$ that returns such an $M$ and works within polynomial time w.r.t. $1/\err$.
  \item There is an algorithm $\probd{\wdt}(m)$, where $m\in\N_0$, that returns the value $\wdt(m)$ and works within polynomial time w.r.t $m$.
\end{enumerate}
\end{definition}

%\begin{lemma}\label{lem:tractable:length}
%Let \wdt be a tractable length distribution. 
%\begin{enumerate}
%	\item There is $k\in\Ns0$ such that       
%	\begin{equation}\label{eq:tractable}
%		\wdt(\N^{>(\log\frac{1}{\err})^k})\le\err,\quad\hbox{for all $\err\in(0,1)$}.
%	\end{equation}
%	\item
%    The expected length of \lbdu\wdt is finite.
%\end{enumerate}
%\end{lemma}
%\begin{proof}
%    The first statement follows when we note that, as \wdt is tractable, there is $k\in\Ns0$ such that, for any $\err\in(0,1)$, there is $M\le\big(\log(1/\err)\big)^k$ such that $\wdt(\N^{>M})\le\err$.
%	For the second statement, we have the following about the expected length of \lbdu\wdt.
%	\begin{equation}\label{eq:lem:tractable}
%	\ev{\dlen{\lbdu{T}}}=\ev{T}=\sum_{i\in\N_0}i\wdt(i)\le\sum_{i\in\N_0}i\wdt(\N^{>i-1}).
%	\end{equation}
%	Using $\err=1/2^{(i-1)^{1/k}}$ in~\eqref{eq:tractable}, we have that  $\wdt(\N^{>i-1})\le1/2^{(i-1)^{1/k}}$, which implies that the series in~\eqref{eq:lem:tractable} is finite.
%\end{proof}
%
%
\begin{figure}[ht]
\begin{center}
\parbox{0.48\textwidth}
{
\hspace*{0.3\algoindent} %Algorithm 
$\ualgo_{\wdt}(\aut, \err)$   \label{alg:univ:approx}
\pssn \hspace*{\algoindent}
\err := $\min(\err,1/6)$;
\\ \hspace*{\algoindent}
$n$ := $\big\lceil 5/(\err-5\err^2)^2\big\rceil$;
\\ \hspace*{\algoindent}
$M$ := $\maxlen_{\wdt}(\err^2)$;
\\ \hspace*{\algoindent}
for each $\ell=0,\ldots,M$
\\ \hspace*{1.7\algoindent} 
$t_\ell := \probd{T}(\ell)$;
\\ \hspace*{\algoindent}
$D$ := $\big(t_0,\ldots,t_M,1-\sum_{\ell=0}^Mt_\ell\big)$;
\\ \hspace*{\algoindent}
i := 0;
\\ \hspace*{\algoindent}
while (i $<$ $n$):
\\\hspace*{1.7\algoindent} 
$\ell$ := $\selectfin(D)$;
\\\hspace*{1.7\algoindent} 
if ($\ell\not=\none$) $w$ := $\uselect(s,\ell)$;
\\ \hspace*{1.7\algoindent} 
i := i+1;
\\ \hspace*{1.7\algoindent}
if ($\ell\not=\none$ and $w\notin\lang{\aut}$) 
\\ \hspace*{2.3\algoindent} 
return \false;
\\ \hspace*{\algoindent} 
return \true;
\\ \hspace*{\algoindent} 
}
\quad
\parbox{0.48\textwidth}
{
\hspace*{0.3\algoindent} %Algorithm 
$\ualgo_{\wdt}(\aut, \err)$   \label{alg:univ:approx}
\pssn \hspace*{\algoindent}
\err := $\min(\err,1/6)$;
\\ \hspace*{\algoindent}
$n$ := $\big\lceil 5/(\err-5\err^2)^2\big\rceil$;
\\ \hspace*{\algoindent}
$M$ := $\maxlen_{\wdt}(\err^2)$;
\\ \hspace*{\algoindent}
for each $\ell=0,\ldots,M$
\\ \hspace*{1.7\algoindent} 
$t_\ell := \probd{T}(\ell)$;
\\ \hspace*{\algoindent}
$D$ := $\big(t_0,\ldots,t_M,1-\sum_{\ell=0}^Mt_\ell\big)$;
\\ \hspace*{\algoindent}
i := 0;\quad cnt := 0;
\\ \hspace*{\algoindent}
while (i $<$ $n$):
\\\hspace*{1.7\algoindent} 
$\ell$ := $\selectfin(D)$;
\\\hspace*{1.7\algoindent} 
if ($\ell\not=\none$) $w$ := $\uselect(s,\ell)$;
\\ \hspace*{1.7\algoindent} 
i := i+1;
\\ \hspace*{1.7\algoindent}
if ($\ell=\none$ or $w\in\lang{\aut}$) 
\\ \hspace*{2.3\algoindent} 
cnt := cnt+1;
\\ \hspace*{\algoindent} 
if (cnt $<n$) return \false 
\\ \hspace*{\algoindent} 
else return \true;
}
\parbox{0.9\textwidth}{\caption{On the left is the PRAX for NFA universality with respect to a certain tractable word distribution \lbdu\wdt---see Theorem~\ref{th:univ:approx}. The value 1/6 in $\min(\err,1/6)$ can be replaced with any value $<1/5$. The version of the algorithm on the right is logically equivalent; it mimics the process in Fig.~\ref{fig:param:est2} and is intended to give a more clear explanation of correctness. }\label{fig:univ:approx}}
\end{center}
\end{figure}

\begin{theorem}\label{th:univ:approx}
	Let $\wdt$ be a tractable word distribution. Algorithm $\ualgo_{\wdt}(\aut, \err)$ in Fig.~\ref{fig:univ:approx} is a polynomial randomized approximation algorithm for NFA universality relative to \lbdu\wdt. 
\end{theorem}
\begin{proof}
For brevity we use  $A(\aut,\err)$ to refer to $\ualgo_{\wdt}(\aut,\err)$. The algorithm needs to be able to select repeatedly either a word $w$ of length $\le M$  from $\lbdu T$ or the outcome `$\none$'. The finite probability distribution $D$ refers to the outcomes $\{0,1,\ldots,M,\none\}$; that is, a length $\ell\le M$ or `$\none$'. Statement $w\select W^M$ of the process in Fig.~\ref{fig:param:est2} corresponds, for $W=\lbdu T$,  to  the first two statements of the while loop: First, select $\ell$ to be either a length $\le M$ or `$\none$' using $\selectfin(D)$ of Lemma~\ref{lem:selectfin}. If a length $\ell$ is selected then use $\uselect(s,\ell)$ to get a word from $\al_s^\ell$.
\pnsi
Next we need to verify the three conditions about $A(\aut,\err)$ in Definition~\ref{D:PRAA}. For the \emshort{first} one, suppose that \aut is universal with respect to $\dom\lbdu\wdt$, that is, $\lbdu{\wdt}(\aut)=1$, equivalently $\lang\aut = \dom\lbdu\wdt$. Then, every selection $w$ from $\lbdu\wdt^M$ is either \none or a word in $\lang\aut$, so the algorithm will return \true. For the \emshort{second} condition, we assume that $\lbdu\wdt(\aut)<1-\err$. As \wdt is tractable and $M=\maxlen_\wdt(\err^2)$, we have that $\wdt(\N^{>M})\le\err^2$. Consider the version of the algorithm $A(\aut,\err)$ on the right and the random process in Fig.~\ref{fig:param:est2} and assume that it selects exactly the same words $w$ as $A(\aut,\err)$ does. Then, algorithm $A(\aut,\err)$ returns \true if and only if the random variable \cnt in Lemma~\ref{lem:univ:maxlen:estim} takes the value $n$. Let $x=5$.  Then, using $\upar=1$ and $\uparg=1-\err$ in Lemma~\ref{lem:univ:maxlen:estim}, we have $(\upar-\uparg)/\wdt(\als^{>M})\ge\err/\err^{2}=(1/\err)^{2-1}>x$ and
\begin{eqnarray*}
\prob{A(\aut,\err)=\true} &=& \prob{\cnt=n} = \prob{\cnt/n\ge1}
%\\&\le & 
\le\frac{1}{x}\>+\>\frac{1}{4n(\err-x\err^2)^2} \le  \frac 1 4.
\end{eqnarray*}
For the \emshort{third} condition, first note that $n=O(1/\err^2)$. As \wdt is tractable, the magnitude of $M$ and the running times of $\maxlen_\wdt(\err^2)$  and $\probd\wdt(M)$ are polynomially bounded as required. Testing whether $w$ is in $\lang\aut$ can be done in time $O(|w||\aut|)$, which is also polynomially bounded, as $|w|\le M$. Thus, $A(\aut,\err)$ runs within polynomial time w.r.t.  $|\aut|$ and $1/\err$, as required.
\end{proof}

%%%%%%%%%%%%%%%%%%%%%%%%%%%%%%%%%%%%%%%%%%%%%%%%%%%%%%%%%%%%%%%%%%%%%%%%%
\subsection{PRAX for the Lambert and Dirichlet Distributions}\label{sec:cases}
We apply next Theorem~\ref{th:univ:approx} to the Lambert and Dirichlet Distributions.

\begin{corollary}\label{cor:geom}
	There is a polynomial randomized approximation algorithm for NFA universality relative to the Lambert distribution. In fact the algorithm works in time $O\big(|\aut|(1/\err)^2\log(1/\err)\big)$ under the assumption of constant cost of $\tosscoin$ and of arithmetic operations\footnote{If the precision $q$, say, of arithmetic needs to be accounted for then a polynomial in $q$ term would be factored in.}.
\end{corollary}
\begin{proof}
      First we need to show that the Lambert distribution is tractable. We have that
	  $\ldla{s,d}(\N^{>M})\le\err$ when 
	  $$M\ge\log_{s}(1/\err)+d-1$$ 
	  and the smallest such $M$ is of magnitude $O(\log(1/\err))$. Computing the $M$ and each value $\probd{\ldla{s,d}}(\ell)=(1-1/s)(1/s)^{\ell-d}$, for $\ell\ge d$, can be done within polynomial time. Under the assumption of constant costs, computing $M$ has constant cost and computing each $(1-1/s)(1/s)^{\ell-d}$ has cost $O(\ell)$. Hence, the time of the algorithm in Fig.~\ref{fig:univ:approx} is $O\big(M^2+n\times(M+|\aut|M)\big)$, where recall $n=O(1/\err^2)$.
\end{proof}

%For the Dirichlet distribution \wdrd, we shall  assume that the constant $t$ is an integer in $\Ns{>2}$. %

For the case of the Dirichlet distribution we need the following lemma

\begin{lemma}\label{lem:dirichlet}
Let $0<\varepsilon<1$ and let $t>1$.
Then for $M\ge \sqrt[t-1]{\frac{1}{\varepsilon}}$ we have
$$\frac{1}{\zeta(t)}\sum_{n=M+1}^\infty \frac{1}{n^t}<\varepsilon,$$ and for $M<-1+\sqrt[t-1]{\frac{1}{t\varepsilon}}$ we have that
$$\frac{1}{\zeta(t)}\sum_{n=M+1}^\infty \frac{1}{n^t}\ge\varepsilon.
$$
\end{lemma}
\begin{proof}
We use the well known fact that for $a>0$ and $t>1$ we have $\int_a^\infty \frac{1}{x^t} dx =  \frac{1}{(t-1)a^{t-1}}$ and the Integral Test.
By the Integral Test we have that 
$$
\frac{1}{t-1}=\int_{1}^\infty \frac{1}{x^t} dx \le \zeta(t) \le 1+\int_{1}^\infty \frac{1}{x^t} dx=\frac{t}{t-1},  
$$
and for $M\ge 1$ we have
$$
\frac{1}{(t-1)(M+1)^{t-1}}=\int_{M+1}^\infty \frac{1}{x^t} dx \le \sum_{n=M+1}^\infty \frac{1}{n^t} \le \int_{M}^\infty \frac{1}{x^t} dx = \frac{1}{(t-1)M^{t-1}}.
$$
Hence for $M\ge \sqrt[t-1]{\frac{1}{\varepsilon}}$ we have
\begin{eqnarray*}
\sum_{n=M+1}^\infty \frac{1}{n^t} &\le & \frac{1}{(t-1)M^{t-1}} \\
&\le & \frac{1}{t-1} \cdot \varepsilon <\zeta(t)\varepsilon. 
\end{eqnarray*}
On the other hand, for $M<-1+\sqrt[t-1]{\frac{1}{t\varepsilon}}$ we get
\begin{eqnarray*}
\sum_{n=M+1}^\infty \frac{1}{n^t} &\ge & \frac{1}{(t-1)(M+1)^{t-1}} \\
&\ge & \frac{t}{t-1} \cdot \varepsilon > \zeta(t)\varepsilon.
\end{eqnarray*}
\end{proof}

\begin{corollary}\label{cor:riem}
	There is a polynomial randomized approximation algorithm for NFA universality relative to the Dirichlet distribution. In fact the algorithm works in time $O\big(|\aut|(1/\err)^2\sqrt[t-1]{1/\err}\,\big)$ under the assumption of constant cost of $\tosscoin$ and of arithmetic operations.
\end{corollary}
\begin{proof}
      First we need to show that the Dirichlet distribution is tractable. Using Lemma~\ref{lem:dirichlet}, we have that $\lddi{t,d}(\N^{>M})\le\err$ holds true for any integer $M$ such that
      \[
      M\> \ge \> \sqrt[t-1]{\frac{1}{\err}}\>  +d-1,
      \]
	which implies that the above $M$ is of polynomially bounded  magnitude  as required. Computing $M$ and each $\probd{\lddi{t,d}}(\ell)=(1/\zeta(t))(\ell+1-d)^{-t}$, for $\ell\ge d$, can be done within polynomial time. Under the assumption of constant costs, computing $M$ has constant cost and computing each $(1/\zeta(t))(\ell+1-d)^{-t}$ also has constant cost. Hence, the time of the algorithm in Fig.~\ref{fig:univ:approx} is $O\big(M+n\times(M+|\aut|M)\big)$, where recall $n=O(1/\err^2)$.
\end{proof}

\subsection{A PAX for  Universality of Unary NFAs}\label{sec:PAXforUniv}
Using the concept of a tractable distribution $T$, which is assumed fixed,  we define below a  simple PAX for universality of unary NFAs, that is, NFAs over the alphabet $\al_1=\{0\}$.

\begin{figure}[ht]
\begin{center}
\parbox{0.48\textwidth}
{
\hspace*{0.3\algoindent} %Algorithm 
$\ualgounary_{\wdt}(\aut, \err)$   \label{alg:univ:approx2}
\\ \hspace*{\algoindent}
$M$ := $\maxlen_{\wdt}(\err)$;
\\ \hspace*{\algoindent}
for each $\ell=0,\ldots,M$
\\ \hspace*{1.7\algoindent} 
if \big($0^\ell\notin\lang{\aut}$\big) 
return \false;
\\ \hspace*{\algoindent} 
return \true;
\\ \hspace*{\algoindent} 
}
\parbox{0.9\textwidth}{\caption{The algorithm tests whether all words of length up to $M$ are accepted by the given NFA \aut, that is, whether $\al_1^{\le M}\subseteq\lang{\aut}$. If yes then $\lbdu\wdt\big(\al_1^{>M}\big)\le\err$  implies $\lbdu{\wdt}\big(\lang\aut\big)\ge1-\err$, as required. }\label{fig:univ:PAX}}
\end{center}
\end{figure}

\pnsn
We have that the NFA universality problem is NP-complete, and we note that the PAX algorithm could be faster than the PRAX one in Theorem~\ref{th:univ:approx}. Of course the case of unary alphabets normally falls outside the context of coding and information theory so the value of the PAX algorithm is not clear.

%%%%%%%%%%%%%%%%%%%%%%%%%%%%%%%%%%%%%%%%%%%%%%%%%%%%%%%%%
\section{Concluding Remarks}\label{sec:last}
The concept of approximate maximality of a block code introduced in \cite{KMR:2018} leads naturally to the concept of approximately universal  block NFAs and also of approximately universal NFAs in general relative to a desirable probability distribution on words. These concepts are meaningful in coding theory where the languages of interest are finite or even regular and can be represented by automata, \cite{MaRoSi:2001,Vardy:1998,BePeRe:2009,KMR:2018}.  
\pnsi
Algorithm $\ualgo$ can be used to decide approximate universality (relative to tractable distributions) of any context-free language, or even any polynomially decidable language $\lang{\aut}$, where now \aut would be a context-free grammar, or a polynomial Turing machine. Of course universality of context-free grammars (or  Turing machines) is undecidable! However, extending our approach to grammars, or Turing machines, is outside of our motivation from coding and information theory and we cannot tell whether it could lead to any meaningful  results. 
\pnsi
Our approach can possibly be used to address other similar hard problems. 
For example, consider the empty DFA intersection problem \edfai. 
Let $\upar\in[0,1]$. We say that a DFA \aut is \emdef{\upar-empty relative} to a word distribution \wrd, if $\wrd(\aut)\le\upar$. 
For example, a block DFA \autb of word length $\ell$ is \emdef{\upar-empty} relative to the uniform distribution on $\al_s^\ell$, if $\card{\lang{\autb}}/\asz^\ell\le\upar$. 
Let \autac denote the complement of the DFA \aut relative to $W$, that is, the DFA accepting $\dom W\sm\lang{\aut}$. In particular, here we assume that $\dom W=\als^*$. Then, \autac can be constructed from \aut in linear time. 

\begin{remark}\label{rem:empty}
	A  DFA \aut is \upar-empty relative to \wrd if and only if \autac is $(1-\upar)$-universal relative to \wrd.
\end{remark} 
\pnsn
As stated already in \cite{RaShZh:2012}, given DFAs $\aut_1,\ldots,\aut_m$, deciding whether their intersection is empty is equivalent to deciding whether the union of $\autac_1,\ldots,\autac_m$ accepts $\als^*$. Note here that, in linear time, one can compute an NFA \aut accepting that union. The question of whether the intersection of $\autac_1,\ldots,\autac_m$ is \upar-empty (relative to some \wrd) is equivalent to whether the NFA \aut is $(1-\upar)$-universal (relative to \wrd). Thus, Corollary~\ref{th:ubapprox} or Theorem~\ref{th:univ:approx} can be used to give a randomized approximate answer to the \upar-emptiness problem for DFA intersection.  
\pssi
Another hard problem that can possibly be approximated via a tractable distribution $T$ is whether two languages are approximately equal (or two NFAs are approximately equivalent). In analogy to the universality index of a language, here one can define the overlap index of two languages to be the probability that a word selected from $T$ is not in the  symmetric difference of the two languages. 
\pssi
In closing we note that every coNP language $L$ can be expressed as a $[0,1]$-value language $L_{\val}$ and, therefore, it can be approximated by languages $L_{\val,\err}$. However, the study of this generalization is outside the scope of the present paper, so  we leave it as a topic for future research.
\if\DRAFT1\pssi
\sk{Are there special cases where approx is not hard? See if \cite{AFHHJOW:2021} helps. }
\pssi\sk{TO BE FINALIZED}\fi

%%%%%%%%%%%%%%%%%%%%%%%%%%%%%%%%%%%%%%%%%%%%%%%%%%%%%%%%%
%%\bibliographystyle{abbrv}
\bibliographystyle{plain}
\bibliography{ppr}

%%%%%%%%%%%%%%%%%%%%%%%%%%%%%%%%%%%%%%%%%%%%%%%%%%%%%%%%%
\end{document}